# Energy conversion efficiency from a high order soliton to fundamental solitons in presence of Raman scattering

ROBI KORMOKAR,[*] MD HOSNE MOBAROK SHAMIM, AND MARTIN ROCHETTE

*Department of Electrical and Computer Engineering, McGill University, 3480 University Street, Montréal, Québec, H3A 0E9, Canada*
*[*]robi.kormokar@mail.mcgill.ca*

**Abstract:** We formulate the energy conversion efficiency from a high-order soliton to fundamental solitons by including the influence of interpulse Raman scattering in the fission process. The proposed analytical formula agrees closely with numerical results of the generalized nonlinear Schrödinger equation as well as to experimental results, while the resulting formulation significantly alters the energy conversion efficiency predicted by the Raman-independent inverse scattering method. We also calculate the energy conversion efficiency in materials of different Raman gain profiles such as silica, ZBLAN and chalcogenide glasses ($As_2S_3$ and $As_2Se_3$). It is predicted that ZBLAN glass leads to the largest energy conversion efficiency of all four materials. The energy conversion efficiency is a notion of utmost practical interest for the design of wavelength converters and supercontinuum generation systems based on the dynamics of soliton self-frequency shift.

## 1. Introduction

Soliton self-frequency shift (SSFS) is an intrapulse Raman scattering process that is key for wavelength conversion and supercontinuum generation [1-7]. Among the several nonlinear wavelength conversion mechanisms, SSFS is of particular interest because of its wide tunability range beyond tens of THz [8-11], and the absence of phase-matching constraints. SSFS is best observed with short pulses, as SSFS increases proportionally with the fourth power of inverse pulse duration [2, 12]. This is particularly the case when an Nth order soliton or mother soliton (MS) experiences fission and splits into N fundamental solitons, from which solitons of femtosecond duration typically emerge [1, 13]. Exploiting the soliton fission mechanism for SSFS based wavelength conversion has been reported by different groups [14-16]. For example, Gauthier et al. reported the soliton fission of a MS into solitons experiencing SSFS for wavelength conversion up to 4.8 μm in $InF_3$ fiber [14]. Alamgir et al. reported continuously tunable Raman solitons over the frequency range of 2.047-2.667 μm using soliton fission and SSFS in an $As_2S_3$ microwire [16]. Soliton fission is also a fundamental mechanism for supercontinuum generation in fiber with anomalous dispersion [5]. The energy conversion efficiency (ECE) from the MS to each fundamental soliton is of practical interest for wavelength conversion applications. For instance, one may wish to optimize the energy transfer from the MS specifically towards the fundamental soliton that experiences the largest amount of SSFS [15, 16].

Using the inverse scattering method (ISM), Kodama et al. have shown that an Nth order soliton is a bound state of N fundamental solitons, as a solution to the nonlinear Schrödinger equation (NLSE) [13]. They have also shown analytically that the bound state Nth order soliton is breaking, i.e., experiences fission, under sufficient high-order linear and/or nonlinear perturbation in the fiber. According to the ISM, each fundamental soliton after fission carries an energy given by [1, 13, 17]

$$E_k = \frac{(2N + 1 - 2k)}{N^2} E_0, \quad (1)$$

where $N = (\gamma_0 E_0 T_0/2|\beta_2|)^{1/2}$ with $E_0$ and $T_0$ being the energy and duration of the MS, respectively, $\beta_2$ is the group velocity dispersion (GVD) coefficient, and $\gamma_0$ is the waveguide nonlinear parameter at center frequency $\omega_0$. The index $k = 1\ldots N$ labels each soliton emerging from fission. Each soliton thus carries an energy $E_k$ given by Eq. 1, leading to an ECE given by $ECE_{ISM,k} = (100 E_k/E_0)\%$. Of particular interest for wavelength conversion is the $k = 1$ Raman soliton because it is the shortest, the most powerful, and most importantly, it experiences the largest amount of SSFS. We however recently noticed numerically and experimentally that the $k = 1$ Raman soliton emerging from soliton fission often carries significantly more energy than predicted by the ISM, leading to $ECE > ECE_{ISM,1}$ [15].

In this paper, we demonstrate that the ECE evaluated from numerical simulations and from experiment agree together but diverge from the theoretical prediction of $ECE_{ISM,1}$. We subsequently propose an improved analytical formulation of ECE for the $k = 1$ Raman soliton, derived by adding an energy transfer contribution from interpulse Raman scattering in between the $k = 2\ldots N$ solitons to the $k = 1$ soliton. We show that the $ECE_{ISM,1}$ diverges rapidly from numerical predictions and experiment with MS of increasing order and decreasing duration. This study clarifies the dynamics of Raman induced soliton fission, as well as providing a more precise evaluation of ECE, a topic of utmost interest for the design of SSFS based wavelength converters and supercontinuum sources.

## 2. Theory of Enhanced Energy Conversion Efficiency

### 2.1 Generalized Nonlinear Schrödinger Equation

The propagation of an optical pulse in a nonlinear and dispersive optical fiber is well described by the generalized nonlinear Schrödinger equation (GNLSE) [1]

$$\frac{\partial A}{\partial z} + \frac{1}{2}\left(\alpha_0 + \sum_{n=1}^{\infty} \frac{i^n \alpha_n}{n!} \frac{\partial^n}{\partial T^n}\right) A - i \sum_{n=2}^{\infty} \frac{i^n \beta_n}{n!} \frac{\partial^n A}{\partial T^n} = \\ i\left(\gamma_0 + \sum_{n=1}^{\infty} \frac{i^n \gamma_n}{n!} \frac{\partial^n}{\partial T^n}\right)\left(A(z,T) \int_0^{\infty} R(T')|A(z, T - T')|^2 dT'\right), \quad (2)$$

where A is the slowly varying pulse envelope, $\alpha_n = d^n\alpha/d\omega^n$, $\beta_n = d^n\beta/d\omega^n$, $\gamma_n = d^n\gamma/d\omega^n$ are the Taylor coefficients of the frequency dependent fiber loss ($\alpha$), propagation constant ($\beta$) and nonlinear parameter ($\gamma$), respectively, evaluated at the center frequency $\omega_0$. T is time in a frame of reference moving at group velocity of the pulse. R(T) is the nonlinear response function defined as [1]

$$R(T) = (1 - f_R)\delta(T) + f_R h_R(T), \quad (3)$$

where $f_R$ represents the fractional contribution of the delayed Raman response to nonlinear polarization, $\delta(T)$ is the Dirac delta function, and $h_R(T)$ is the Raman response function. Using the single peak Lorentzian model, $h_R(T)$ can be written as

$$h_R(T) = (\tau_1^{-2} + \tau_2^{-2})\tau_1 \exp(-T/\tau_2) \sin(T/\tau_1), \quad (4)$$

where $\Omega_R = 1/\tau_1$ is the angular frequency of the Raman gain peak and $\Gamma_R = 1/\tau_2$ is the bandwidth of the Raman gain response [18].

## 2.2 Soliton Constitution and Fission

By ignoring high-order linear and nonlinear terms and by considering a lossless medium, the NLSE can be written in a normalized form

$$\frac{\partial q}{\partial Z} = \frac{i}{2}\frac{\partial^2 q}{\partial \tau^2} + i|q|^2 q, \quad (5)$$

with

$$q = \frac{N}{\sqrt{P_0}}A, \quad Z = \frac{z}{L_D}, \quad \text{and} \quad \tau = \frac{T}{T_0}, \quad (6)$$

where $P_0$ and $T_0$ are the pulse peak power and duration, and $L_D = T_0^2/|\beta_2|$ is the dispersion length. By using the inverse scattering method developed by Zakharov et al. [19], Satsuma et al. showed that a solution of Eq. (5) is an $N^{th}$ order soliton of the form $q = N\text{sech}(\tau)$, which itself contains N fundamental solitons with amplitude [20]

$$\eta_k = 2(N - k) + 1, \quad k = 1, 2, \ldots N \quad (7)$$

and where fundamental solitons are expressed as [17]

$$q_k(\tau, Z) = \eta_k \text{sech}[\eta_k(\tau + \kappa_k Z - \tau_{0k})] \exp\left[-i\kappa_k \tau + \frac{i}{2}(\eta_k^2 - \kappa_k^2)Z - i\sigma_{0k}\right], \quad (8)$$

where $\kappa_k$ is related with the change in group velocity of the $k^{th}$ soliton component. Satsuma et al. showed that $\kappa_k = 0$ and all fundamental solitons of a high-order soliton travel at the same group velocity, leading to a bound state of solitons [20]. Because of the phase interference among fundamental solitons, the resulting high-order soliton profile and spectrum evolve periodically with a period of $Z = \pi/2$. By using Eq. (6) and (7) in Eq. (8) and by setting the constant phase term $\sigma_{0k}$ and the temporal offset of each soliton $\tau_{0k}$ to zero, the fundamental solitons are expressed as

$$A_k(T, 0) = \frac{(2N + 1 - 2k)}{N}\sqrt{P_0}\,\text{sech}\left[\frac{T}{T_0/(2N + 1 - 2k)}\right]. \quad (9)$$

From Eq. (9), the peak power and duration of each fundamental soliton component is

$$P_k = \frac{(2N + 1 - 2k)^2}{N^2}P_0, \quad (10a)$$

and

$$T_k = \frac{T_0}{(2N + 1 - 2k)}. \quad (10b)$$

A high-order soliton becomes however unstable and eventually fissions under the influence of linear and/or nonlinear perturbations such as Raman effect, third-order dispersion, and self-steepening. As a result, fundamental solitons come apart and propagate with a peak power and pulse duration defined in Eq. (10). According to the ISM, the ECE of each soliton is given by

$$\text{ECE}_{\text{ISM},k} = \frac{(2N + 1 - 2k)}{N^2} \times 100\%. \quad (11)$$

In a glassy medium such as an optical fiber, soliton propagation is also accompanied with intrapulse Raman scattering that gradually transfers energy from short wavelength contents of the pulse to the long wavelengths [2, 12]. As a result, the soliton experiences SSFS following [2]

$$\Omega(z) = -\frac{8T_R |\beta_2|}{15 T_0^4} z_{\text{eff}}, \quad (12)$$

where $\Omega$ is SSFS in rad/s, $T_R$ is Raman time, and

$$z_{\text{eff}} = \frac{1}{4\alpha}(1 - e^{-4\alpha z}), \quad (13)$$

is the effective length of the fiber. Equation (12) shows that SSFS increases proportionally with the effective fiber length and with inverse pulse duration to the fourth order. Since the k = 1 soliton has the shortest duration, according to Eq. (10b), this soliton therefore experiences the largest amount of SSFS compared to the other soliton components.

Figure 1 shows a typical example of Raman induced soliton fission process of a N = 3 MS in a 15 m long lossless standard single mode fiber (SSMF). To demonstrate the concept of Raman induced soliton fission, the simulation only considers the Raman effect as the sole high-order nonlinear effect. Fiber parameters are $\beta_2$ = - 21.61 ps$^2$/km, and $\gamma_0$ = 1.42 W$^{-1}$/km at the initial pulse wavelength of $\lambda_0$ = 1.55 µm. The Raman response function given in Eq. (4) is used for silica with $\tau_1$ = 12.2 fs, $\tau_2$ = 32 fs and $f_R$ = 0.18. The initial pulse duration and peak power are 283 fs and 1.7 kW, respectively. The simulation is performed by numerically solving the GNLSE of Eq. (2) using a split-step Fourier method (details in section 4). As seen in Fig. 1, the k = 1 Raman soliton emerges due to Raman induced soliton fission and experiences the most SSFS. It is also observed that $\text{ECE}_{k=1}$ = 66.0% whereas $\text{ECE}_{\text{ISM},1}$ = 55.6%. The numerical simulations and ISM results thus significantly differ in this example.

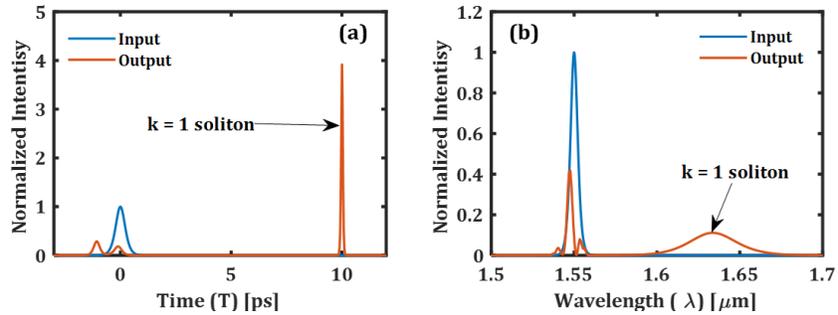

Fig. 1. Simulated (a) temporal and (b) spectral evolution of a third-order MS following a Raman induced soliton fission process. The pulse initially has a center wavelength of $\lambda_0$ = 1.55 µm, pulse duration of 283 fs and peak power of 1.7 kW. The medium is a 15 m long lossless SSMF with $\beta_2$ = - 21.61 ps$^2$/km, $\gamma_0$ = 1.42 W$^{-1}$/km at $\lambda_0$ = 1.55 µm.

The observed ECE enhancement with respect to the ISM prediction is explained from a transfer of energy in between solitons following the fission process, known as interpulse Raman scattering [21, 22], and in contrast with intrapulse Raman scattering. In a bound state, a high-order soliton experiences a periodic evolution due to a mutual interaction between GVD and self-phase modulation (SPM) [17, 20]. High-order linear and/or nonlinear perturbations break down the bound state, which triggers a fission process, and the MS breaks into fundamental soliton components. In any glassy medium, Raman scattering is the dominant mechanism that triggers soliton fission [1]. The Raman effect transfers soliton energy towards lower frequency components via intra- and interpulse Raman scattering. Intrapulse Raman scattering is responsible for SSFS. Interpulse Raman scattering also occurs after the fission event, amplifying solitons at long wavelength from solitons at short wavelengths, leading to energy transfer in between solitons that temporally overlap. Since the k = 1 Raman soliton experiences the most redshift from SSFS, it becomes amplified by the remaining k = 2…N solitons for as long as they temporally overlap. For this reason, the k = 1 Raman soliton is expected to get an enhanced ECE with respect to the ISM prediction. Note that for a MS with N being a non-integer value, soliton fission is followed by fundamental solitons and dispersive waves.

## 3. Analytical Formulation of Enhanced ECE

In addition to the ISM prediction of ECE that occurs during the fission of the MS, a new formulation of ECE should also consider the energy transfer ($\Delta E$) from the k = 2…N solitons to the k = 1 Raman soliton resulting from interpulse Raman scattering that occurs moments after fission. Taking interpulse Raman scattering into account, the new ECE of the k = 1 soliton is expressed as

$$\mathrm{ECE}_{new,1} = \mathrm{ECE}_{ISM,1} + \Delta e, \tag{14}$$

where $\Delta e$ is a correction term to the $\mathrm{ECE}_{ISM,1}$. For interpulse Raman scattering to occur, the temporal overlap in between involved pulses is required. An estimated pump energy from each k = 2…N solitons to the k = 1 Raman soliton is expressed as

$$E_{pk}(z) = \int_{-3T_1+q}^{3T_1+q} P_k \operatorname{sech}^2\left(\frac{T}{T_k}\right) dT, \tag{15}$$

where $T_1$ and $q$ are the pulse duration and temporal position of the k = 1 Raman soliton, respectively. $P_k$ and $T_k$ are the peak power and pulse duration of the k = 2…N solitons. The integration limits in Eq. (15) are set such that when q = 0 the integration provides the energies of the k = 2…N solitons that are overlapped with 99.0% of the total energy of the k = 1 soliton. From Eq. (10) and integration, Eq. (15) transforms into

$$E_{pk}(z) = \frac{E_0(2N + 1 - 2k)}{N^2} \tanh\left[\frac{3(2N + 1 - 2k)}{(2N - 1)}\right] \Lambda_k(q), \tag{16a}$$

where

$$\Lambda_k(q) = \frac{\left(1 - \tanh^2 \frac{q}{T_k}\right)}{\left(1 - \tanh^2 \frac{3T_1}{T_k} \tanh^2 \frac{q}{T_k}\right)}. \tag{16b}$$

The temporal position of the k = 1 soliton is evaluated from the moment method [23]

$$q(z) = \frac{4T_R|\beta_2|^2}{15T_1^4} z^2. \tag{17}$$

Therefore, the total pump energy from the k = 2…N soliton components sums up into

$$E_{pT}(z) = \sum_{k=2}^{N} \frac{E_0(2N+1-2k)}{N^2} \tanh\left[\frac{3(2N+1-2k)}{(2N-1)}\right] \Lambda_k(q). \tag{18}$$

The ECE enhancement rate with propagation distance is proportional to $E_{pT}(z)$. Therefore,

$$\frac{de}{dz} \propto \sum_{k=2}^{N} \left(\frac{(2N+1-2k)}{N^2} \tanh\left[\frac{3(2N+1-2k)}{(2N-1)}\right]\right) \Lambda_k(q). \tag{19}$$

This rate also depends on the effective Raman gain experienced by the $k = 1$ Raman soliton, quantified as

$$\frac{de}{dz} \propto \gamma_0 P_1 f_R \int_{-\infty}^{0} \text{Im}[\tilde{h}_R(\omega)] \text{sech}^2\left[\frac{\pi(\omega - \Omega(z))T_0}{2(2N-1)}\right] d\omega, \tag{20}$$

where $P_1$ is the peak power of the $k = 1$ Raman soliton, and $\text{Im}[\tilde{h}_R(\omega)]$ is the Raman gain function [1]. $\Omega(z)$ is the SSFS experienced by the $k = 1$ Raman soliton. By combining Eqs. (19) and (20), the ECE enhancement rate becomes

$$\frac{de}{dz} \propto \sum_{k=2}^{N} \left(\frac{(2N+1-2k)}{N^2} \tanh\left[\frac{3(2N+1-2k)}{(2N-1)}\right]\right) \Lambda_k(q) \\ \times \gamma_0 P_1 f_R \int_{-\infty}^{0} \text{Im}[\tilde{h}_R(\omega)] \text{sech}^2\left[\frac{\pi(\omega - \Omega(z))T_0}{2(2N-1)}\right] d\omega. \tag{21}$$

By integration of Eq. (21) with respect to z, the correction term in Eq. (14) becomes

$$\Delta e \propto \sum_{k=2}^{N} I_k \left(\frac{(2N+1-2k)}{N^2} \tanh\left[\frac{3(2N+1-2k)}{(2N-1)}\right]\right), \tag{22}$$

where

$$I_k(z) = \int_{0}^{z} \Lambda_k(q) \gamma_0 P_1 f_R \int_{-\infty}^{0} \left[\text{Im}[\tilde{h}_R(\omega)] \text{sech}^2\left[\frac{\pi(\omega - \Omega(z))T_0}{2(2N-1)}\right]\right] d\omega \, dz. \tag{23}$$

From Eq. (14) and Eq. (22), the correction to the ECE predicted by the ISM becomes

$$ECE_{new,1} = ECE_{ISM,1}\left[1 + \kappa \sum_{k=2}^{N} I_k \frac{(2N+1-2k)}{(2N-1)} \tanh\left(\frac{3(2N+1-2k)}{2N-1}\right)\right], \tag{24}$$

including a material-dependent proportionality constant

$$\kappa = \frac{10}{\int_{-\infty}^{0} \text{Im}[\tilde{h}_R(\omega)] d\omega} \tag{25}$$

where the factor of 10 was calculated numerically to fit the analytical results with numerical results. This factor arises from the nonlinear superposition of the fundamental soliton components, not considered in Eq. (19) for the sake of simplicity of the analytical formulation. $ECE_{new,1}$ is an analytical expression of ECE that includes the combined effects of interpulse and intrapulse Raman scattering. The formula of Eq. (24) is therefore a great tool to guide towards the design of SSFS-based wavelength conversion systems, without requiring heavy computing approaches.

## 4. Numerical Validation

The split-step Fourier method (SSFM) is a numerical method which is extensively used for solving pulse propagation in optical fibers [1]. A recent study by Farag et al. has shown that the SSFM provides least cumulative error with fastest computational speed compared to other commonly used numerical methods [24]. For our study, a numerical GNLSE solver is developed to implement the SSFM and solve the GNLSE in Eq. (2) for pulse evolution, using MATLAB. To obtain a more accurate simulation output, a full form of the Raman response function is used instead of using the Raman time constant ($T_R$). The convolution integral in Eq. (2) is calculated using [25, 26]

$$\int_0^\infty R(T')|A(z, T - T')|^2 dT' = (1 - f_R)|A(z,T)|^2 + f_R \Delta T \mathcal{F}^{-1}\big( \mathcal{F}(h_R) \, \mathcal{F}(|A(z,T)|^2)\big), \quad (26)$$

where $\mathcal{F}$ and $\mathcal{F}^{-1}$ represent the Fourier and inverse Fourier transform operations, respectively, and $\Delta T$ is the temporal grid spacing. Several parameters such as temporal window, temporal grid spacing, and step size are carefully set in the GNLSE solver to reach a valid solution. Setting up the proper parameters is essential, especially for modelling MS fission in an optical fiber, because in a soliton fission process the generated $k = 1$ Raman soliton is significantly shorter than the MS, as well as pulses experience large temporal delay and frequency shift, simultaneously. The temporal window is set such that it contains the maximum temporal delay of $k = 1$ Raman soliton over the propagation distance of interest. The temporal grid is sampled densely enough to ensure that the $k = 1$ Raman soliton remains defined by several temporal and spectral points. An adaptive step-size is used to limit the nonlinear phase shift increment in each propagation step during the simulation.

The GNLSE solver is validated by reproducing temporal and spectral results from literature [1, 26]. The first simulation consists into the propagation of a second-order soliton in presence of Raman scattering to simulate SSFS from soliton fission. Fig. 2a shows the generated spectra which matches closely with Fig. 5.24 in Ref. 1. The second simulation case consists into the propagation of a femtosecond MS of order ~8.65 before turning into a supercontinuum. In this simulation, dispersion coefficients up to $10^{th}$ order and nonlinear parameter up to $1^{st}$ order are considered as in Ref. 26. Fig. 2b shows the supercontinuum spectrum which matches closely with Fig. 4 in Ref. 26. In both cases, the GNLSE solver reproduces the results from the literature with high accuracy and supports its validity.

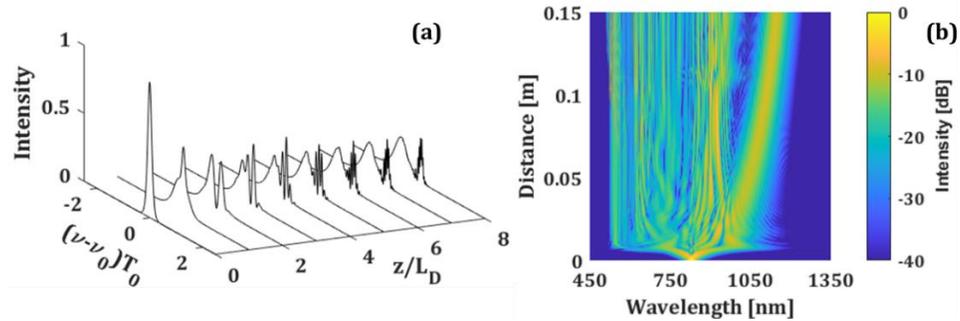

Fig. 2. Simulation results from the GNLSE solver, comparing with results in literature. (a) Spectral evolution of a second-order soliton depicting soliton fission and SSFS and (b) supercontinuum generation in the femtosecond regime.

Figure 3 shows the ECE results of a k = 1 Raman soliton with different initial pulse durations and soliton orders obtained from $ECE_{new}$, the numerical simulations of Eq. (2) ($ECE_{GNLSE}$), and $ECE_{ISM}$. The fiber considered is a lossless SSMF with $\beta_2$ = - 21.61 ps$^2$/km and $\gamma_0$ = 1.42 W$^{-1}$/km at the initial pulse wavelength of $\lambda_0$ = 1.55 μm, and $\tau_1$ = 12.2 fs, $\tau_2$ = 32 fs, and $f_R$ = 0.18 [18, 27].

From Fig. 3, it is observed that the $ECE_{ISM}$ is independent of pulse duration. This is in contrast with the $ECE_{new}$ and $ECE_{GNLSE}$ that diverge gradually from the ISM predictions as the soliton order increases and pulse duration decreases. The ISM however provides asymptotic results to the $ECE_{new}$ and $ECE_{GNLSE}$ as the pulse duration increases. It is also observed that the $ECE_{new}$ and $ECE_{GNLSE}$ agree well except for high soliton orders with short pulse durations, suggesting that another mechanism of energy transfer is triggered. This discrepancy could arise at a point where the k = 2 soliton would begin to benefit of an enhanced ECE by interpulse Raman scattering from solitons k > 2. This is expected to arise when the MS order is sufficiently high and duration is short, such that the spectral overlap between the k = 2 soliton and the Raman gain function significantly increase. For example, with a MS of N = 6 and $T_{FWHM}$ = 700 fs, the $ECE_{k=1}$ = 53.0% and $ECE_{k=2}$ = 22.9%. whereas for a MS with N = 6 and $T_{FWHM}$ = 400 fs, the $ECE_{k=1}$ = 50.0% and $ECE_{k=2}$ = 27.2%. The ISM predicts ECE = 25.0% for the k = 2 soliton. Clearly, the k = 2 soliton eventually benefits of an increased ECE above the expected ISM as the MS gets shorter. This is however an assumption that requires further investigation before being validated.

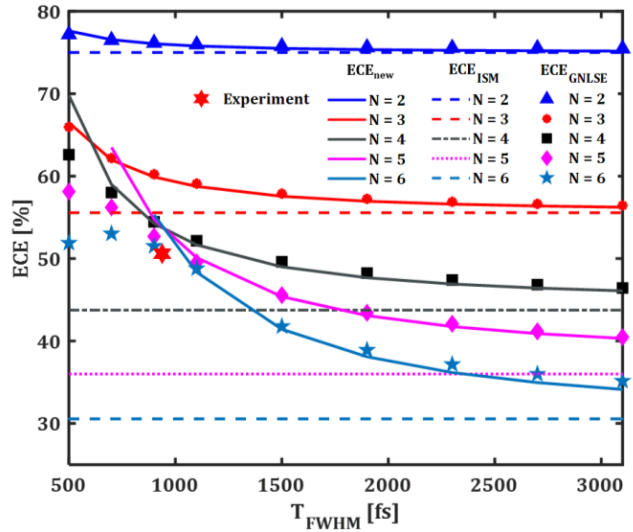

Fig. 3. Energy conversion efficiency of k = 1 soliton after fission of high-order solitons as a function of duration and order.

Figure 4 shows the influence of $\beta_2$ and $\gamma_0$ in the $ECE_{GNLSE}$ of a k = 1 Raman soliton generated from the fission of a MS of N = 3. In the first case, different values of $\beta_2$ are used to observe its effect on the ECE. In the second case, different values of $\gamma_0$ are used to see its effects on the ECE. In both cases, only the peak powers are adjusted to maintain a constant soliton order. From Figs. 4(a) and 4(b) it is observed that the ECE is independent of $\beta_2$ and $\gamma_0$ and their spectral derivative but depend on the MS order and duration.

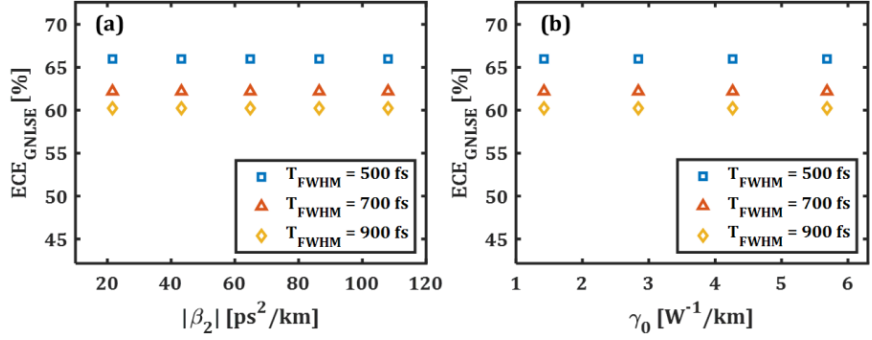

Fig. 4. ECE$_{GNLSE}$ for the fission generated k = 1 Raman soliton from an MS of N = 3, with variable pulse durations propagating in fibers. In (a), variable values of GVD coefficient ($\beta_2$) when $\gamma_0$ = 1.42 W$^{-1}$/km. In (b), variable values of nonlinear parameter ($\gamma_0$) when $\beta_2$ = - 21.61 ps$^2$/km.

## 5. Experimental Validation

Figure 5 shows a schematic of the experimental setup used to probe the ECE of a k = 1 Raman soliton after soliton fission. A mode-locked fiber laser (Calmar-FPL-02CFFPM) generates seed pulses centered at $\lambda_0$ = 1.55 μm with a duration of 295 fs, an average power of 2.8 mW, and at the repetition rate of 20 MHz. An in-line polarization controller is used to maximize the SSFS while an EDFA (Pritel-FA-22-IO) amplifies the seed pulse to an average power of 35.4 mW. The EDFA also broadens the pulse due to normal dispersion of 35 fs/nm. The amplified pulse then turns into a MS that experiences soliton fission and SSFS within a 3 m long SSMF patch cord that leads to a 1% tap coupler ($C_1$). An optical spectrum analyzer (OSA1, Yokogawa-AQ6376) at the 1% port of the coupler $C_1$ probes the initial SSFS of the k = 1 Raman soliton. A variable optical attenuator (VOA, Agilent-81577A) after $C_1$ finely controls the energy of post-fission fundamental soliton before propagation in the nonlinear fiber. The VOA is used as the variable element of this experiment to monitor the spectrum of the soliton that experiences the largest amount of SSFS (i.e., the k = 1 Raman soliton). The nonlinear fiber is a 100 m long SSMF (Thorlabs SMF-28-100), where the soliton experiences SSFS. At the output of the nonlinear fiber, a coupler ($C_2$) splits the signal into two ports for power and spectrum monitoring. The power is measured by an optical power meter (Newport-843-R) while the spectrum is recorded with OSA2 (Yokogawa-AQ6375). For simulation purpose, the wavelength dependence of the VOA and couplers $C_1$ and $C_2$ are measured in the wavelength range of 1.5-1.8 μm with a broadband source. A constant fiber loss of $\alpha_0$ = 0.2 dB/km within the wavelength range of interest is considered, as measured using a broadband source.

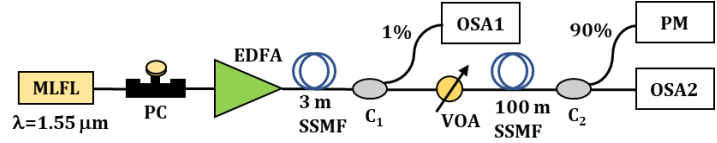

Fig. 5. Schematic of the experimental setup. MLFL: mode-locked fiber laser; PC: polarization controller; EDFA: erbium-doped fiber amplifier; OSA: optical spectrum analyzer; VOA: variable optical attenuator; SSMF: standard single-mode fiber; PM: power meter.

At the output of the EDFA, the MS has a pulse duration of 938 fs and N = 5.57 as estimated from comparison of GNLSE solver and experiment. Figure 6 shows the spectrum measured in OSA1 of the MS generated from the EDFA. The initial Raman shift of 74 nm within the 3 m long SSMF between the EDFA and $C_1$ confirms soliton fission and formation of a Raman soliton. Figure 7 shows the obtained spectra as a function of attenuation levels of the VOA, as well as simulated spectra obtained from numerical evaluation of Eq. (2). Of particular interest is the spectrum of the k = 1 Raman soliton at long wavelengths and experiencing a large variation of SSFS as a function of attenuation levels. In the numerical simulation, GVD coefficients up to $\beta_5$ and nonlinear coefficient up to $\gamma_3$ are used as provided in Table 1.

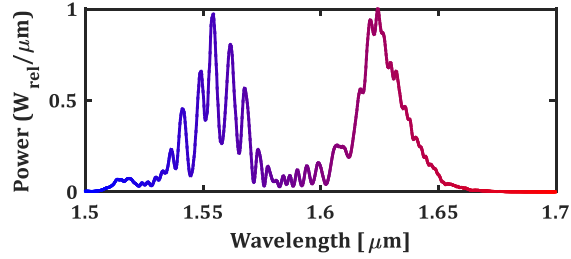

Fig. 6. Soliton fission and initial SSFS observed in OSA1.

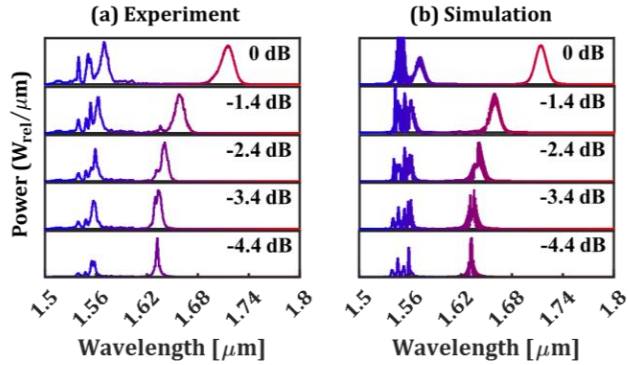

Fig. 7. (a) Measured output (b) simulated spectra of SSFS with increasing VOA levels.

**Table 1. Taylor coefficients of GVD and nonlinear parameter of SSMF fiber.**

| | GVD | | Nonlinear parameter |
|---|---|---|---|
| $\beta_2$: | -21.61 ps$^2$/km | $\gamma_0$: | 1.42 W$^{-1}$/km |
| $\beta_3$: | 0.125 ps$^3$/km | $\gamma_1$: | 0.0029 W$^{-1}$ps/km |
| $\beta_4$: | -3.99 × 10$^{-4}$ ps$^4$/km | $\gamma_2$: | -1.43 × 10$^{-6}$ W$^{-1}$ps$^2$/km |
| $\beta_5$: | 1.99 × 10$^{-6}$ ps$^5$/km | $\gamma_3$: | -1.11 × 10$^{-8}$ W$^{-1}$ps$^3$/km |

The good agreement between numerical simulations and experimental results validates the GNLSE solver results as well as results presented in Fig. 3. The ECE obtained for the k = 1 soliton from the experiment is 50.6%, while ECE$_{GNLSE}$ = 51.1% and ECE$_{new}$ = 53.7%, showing a good agreement in between those values. In contrast, ECE$_{ISM}$ = 32.7%. The results from the GNLSE solver and the experiments clearly shows an enhanced ECE for the k = 1 Raman soliton due to interpulse Raman scattering.

## 6. Effect of Raman Gain Function

According to Eq. (20), the overlap of the spectrum of the k = 1 Raman soliton and the Raman gain function determines the ECE enhancement of the k = 1 Raman soliton. To observe the effects of the Raman gain profile on the ECE of k = 1 Raman soliton, numerical simulations are performed with four different kinds of fiber materials: Silica, ZBLAN, and chalcogenides ($AS_2S_3$ and $As_2Se_3$). The Raman functions of silica and chalcogenides are calculated from Eq. (4). For $As_2S_3$, the Raman response function uses $\tau_1 = 15.5$ fs, $\tau_2 = 230.5$ fs, and $f_R = 0.1$ [16]. For $As_2Se_3$, $\tau_1 = 23.1$ fs, $\tau_2 = 195$ fs [28], and $f_R = 0.1$. An intermediate broadening model is used to generate the Raman response function of ZBLAN [29]. According to this model, the Raman response function of ZBLAN is [30]

$$h_R(T) = \sum_{i=1}^{8} A_i \exp(-\gamma_i T) \exp\left(-\frac{\Gamma_i^2 T^2}{4}\right) \sin(\omega_{v,i}) T, \quad (27)$$

where the parameters $A_i$, $\gamma_i$, $\Gamma_i$, and $\omega_{v,i}$ are tabulated in Ref. 30. Figure 8 shows the Raman gain functions of silica, ZBLAN, and chalcogenides ($As_2S_3$ and $As_2Se_3$) calculated from their Raman responses $h_R(T)$. The inset of Fig. 8 also presents the Raman time ($T_R$), the maximum Raman gain coefficient at 1550 nm and the $\kappa$ parameter for the four materials.

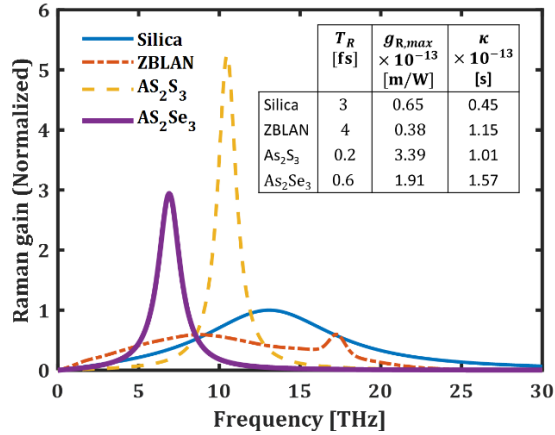

Fig. 8. Raman gain functions of Silica, ZBLAN and chalcogenides ($As_2S_3$ and $As_2Se_3$) normalized with respect to the peak Raman gain of silica. The inset tabulates the Raman time and maximum Raman gain coefficient at 1550 nm, and $\kappa$ parameter of the materials.

Figure 9 shows the $ECE_{GNLSE}$ and the $ECE_{ISM}$ of the k = 1 Raman soliton generated from the fission of a N = 3 MS in the four fiber materials. From Fig. 9 it is observed that for pulse durations of the MS larger than 500 fs, the generated k = 1 Raman solitons obtain maximum amount of ECE in ZBLAN fiber compared to the other three materials. In contrast, the k = 1 Raman solion gets the minimum amount of ECE in $As_2S_3$ fiber. This enhanced ECE in ZBLAN glass is explained by the maximum Raman gain slope of $T_R = 4$ fs, in contrast with the $As_2S_3$ glass that has the smallest Raman gain slope of $T_R = 0.2$ fs, as shown in Fig. 8. As a result, the k = 1 Raman soliton experiences more Raman gain in ZBLAN fiber near the pump frequency than in the other three materials. However, if the pulse duration is short enough (< 1 ps) that the broad spectrum of the k = 1 Raman soliton falls within the gain peak of $As_2S_3$, the ECE starts to enhance as can be seen from Fig. 9. From Figs. 8 and 9, it is therefore evident that the impact of the Raman gain profile is significant on the ECE enhancement of the k = 1 Raman soliton. Figure 9 also suggests that ZBLAN is the best candidate for obtaining enhanced ECE of the k = 1 Raman soliton.

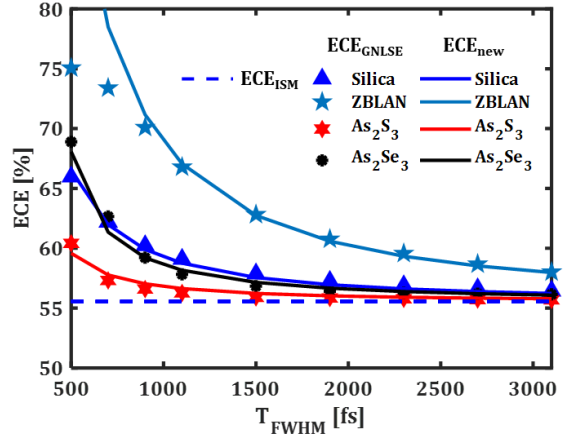

Fig. 9. Energy conversion efficiency of the k = 1 Raman soliton after fission of an N = 3 MS of variable pulse durations and into fibers made of silica, ZBLAN, and chalcogenide glasses.

## 7. Conclusion

In summary, we have demonstrated an enhanced energy conversion efficiency of the k = 1 Raman soliton in a Raman induced soliton fission process. Using the interpulse Raman scattering model, an analytical formula for the enhanced ECE is derived which provides a good fit with the numerical simulation results. The GNLSE solver results reveal that the ECE of k = 1 Raman soliton diverges rapidly from the ISM prediction with increasing soliton order and decreasing pulse duration. The experimental results agree well with the numerical simulation, which proves the existence of enhanced ECE for k = 1 Raman soliton from the fission process triggered by Raman scattering. Finally, ECE results of the k = 1 Raman soliton in four different kinds of fibers show the impact of the Raman gain profile on the ECE enhancement and suggest that ZBLAN is the best material for wavelength conversion using soliton fission. This study is useful for understanding the physical process behind the enhancement of ECE of the k = 1 Raman soliton in a soliton fission process. Moreover, it is useful for the design of optical devices making use of soliton fission, such as SSFS based wavelength converters and supercontinuum sources.

**Data Availability.** Data underlying the results presented in this paper are not publicly available at this time but may be obtained from the authors upon reasonable request.

**Funding.** Natural Sciences and Engineering Research Council of Canada.

**Disclosures.** The authors declare no conflicts of interest.